\documentclass[twocolumn]{aastex631}

\newcommand{\oii}{\hbox{[O$\,${\scriptsize II}]}}
\newcommand{\oiii}{\hbox{[O$\,${\scriptsize III}]}}

\newcommand{\civ}{\hbox{C$\,${\scriptsize IV}}}

\newcommand{\ha}{\hbox{H$\alpha$}}
\newcommand{\hb}{\hbox{H$\beta$}}

\newcommand{\kms}{km\,s$^{-1}$}

\newcommand{\ergs}{erg s$^{-1}$}

\newcommand\jwst{\emph{JWST}}
\newcommand\hst{\emph{HST}}

\shorttitle{J1652}
\shortauthors{Wylezalek al.}
\graphicspath{{./}{figures/}}

\begin{document}

\title{First results from the JWST Early Release Science Program Q3D: \\ Turbulent times in the life of a $z \sim 3$ extremely red quasar revealed by NIRSpec IFU}

\author[0000-0003-2212-6045]{Dominika Wylezalek}
\affiliation{Zentrum für Astronomie der Universität Heidelberg, Astronomisches Rechen-Institut, Mönchhofstr 12-14, D-69120 Heidelberg, Germany}

\author[0000-0002-0710-3729]{Andrey Vayner}
\affiliation{Department of Physics and Astronomy, Bloomberg Center, Johns Hopkins University, Baltimore, MD 21218, USA}

\author[0000-0002-1608-7564]{David S. N. Rupke}
\affiliation{Department of Physics, Rhodes College, Memphis, TN 38112, USA}

\author[0000-0001-6100-6869]{Nadia L. Zakamska}
\affiliation{Department of Physics and Astronomy, Bloomberg Center, Johns Hopkins University, Baltimore, MD 21218, USA}
\affiliation{Institute for Advanced Study, Princeton, NJ 08540, USA}

\author[0000-0002-3158-6820]{Sylvain Veilleux}
\affiliation{Department of Astronomy and Joint Space-Science Institute, University of Maryland, College Park, MD 20742, USA}

\author[0000-0001-7572-5231]{Yuzo Ishikawa}
\affiliation{Department of Physics and Astronomy, Bloomberg Center, Johns Hopkins University, Baltimore, MD 21218, USA}

\author[0000-0002-6948-1485]{Caroline Bertemes}
\affiliation{Zentrum für Astronomie der Universität Heidelberg, Astronomisches Rechen-Institut, Mönchhofstr 12-14, D-69120 Heidelberg, Germany}

\author[0000-0003-3762-7344]{Weizhe Liu}
\affiliation{Department of Astronomy and Joint Space-Science Institute, University of Maryland, College Park, MD 20742, USA}

\author[0000-0003-2405-7258]{Jorge K. Barrera-Ballesteros}
\affiliation{Instituto de Astronomía, Universidad Nacional Autónoma de México, AP 70-264, CDMX 04510, Mexico}

\author[0000-0001-8813-4182]{Hsiao-Wen Chen}
\affiliation{Department of Astronomy \& Astrophysics, The University of Chicago, 5640 South Ellis Avenue, Chicago, IL 60637, USA}

\author[0000-0003-4700-663X]{Andy D. Goulding}
\affiliation{Department of Astrophysical Sciences, Princeton University, 4 Ivy Lane, Princeton, NJ 08544, USA}

\author[0000-0002-5612-3427]{Jenny E. Greene}
\affiliation{Department of Astrophysical Sciences, Princeton University, 4 Ivy Lane, Princeton, NJ 08544, USA}

\author[0000-0003-4565-8239]{Kevin N. Hainline}
\affiliation{Steward Observatory, University of Arizona, 933 North Cherry Avenue, Tucson, AZ 85721, USA}

\author{Fred Hamann}
\affiliation{Department of Physics \& Astronomy, University of California, Riverside, CA 92521, USA}

\author[0000-0001-8813-4182]{Timothy Heckman}
\affiliation{Department of Physics and Astronomy, Bloomberg Center, Johns Hopkins University, Baltimore, MD 21218, USA}

\author[0000-0001-9487-8583]{Sean D. Johnson}
\affiliation{Department of Astronomy, University of Michigan, Ann Arbor, MI 48109, USA}

\author[0000-0003-0291-9582]{Dieter Lutz}
\affiliation{Max-Planck-Institut für Extraterrestrische Physik, Giessenbachstrasse 1, D-85748 Garching, Germany}

\author[0000-0001-6126-5238]{Nora Lützgendorf}
\affiliation{European Space Agency, Space Telescope Science Institute, Baltimore, Maryland, USA}

\author[0000-0002-1047-9583]{Vincenzo Mainieri}
\affiliation{European Southern Observatory, Karl-Schwarzschild-Straße 2, D-85748 Garching bei München, Germany}

\author[0000-0002-4985-3819]{Roberto Maiolino}
\affiliation{Kavli Institute for Cosmology, University of Cambridge, Cambridge CB3 0HE, UK; Cavendish Laboratory, University of Cambridge, Cambridge CB3 0HE, UK}

\author[0000-0001-5783-6544]{Nicole P. H. Nesvadba}
\affiliation{Université de la Côte d'Azur, Observatoire de la Côte d'Azur, CNRS, Laboratoire Lagrange, Bd de l'Observatoire, CS 34229, Nice cedex 4 F-06304, France}

\author[0000-0002-3471-981X]{Patrick Ogle}
\affiliation{Space Telescope Science Institute, 3700, San Martin Drive, Baltimore, MD 21218, USA}

\author[0000-0002-0018-3666]{Eckhard Sturm}
\affiliation{Max-Planck-Institut für Extraterrestrische Physik, Giessenbachstrasse 1, D-85748 Garching, Germany}



\begin{abstract}
Extremely red quasars, with bolometric luminosities exceeding $10^{47}$ erg s$^{-1}$, are a fascinating high-redshift population that is absent in the local universe. They are the best candidates for supermassive black holes accreting at rates at or above the Eddington limit, and they are associated with the most rapid and powerful outflows of ionized gas known to date. They are also hosted by massive galaxies. Here we present the first integral field unit (IFU) observations  of a high-redshift quasar obtained by the Near Infrared Spectrograph (NIRSpec) on board the {\it James Webb Space Telescope} (\jwst), which targeted SDSS~J165202.64+172852.3, an extremely red quasar at $z=2.94$. \jwst\ observations reveal extended ionized gas -- as traced by \oiii$\lambda$5007\AA -- in the host galaxy of the quasar, its outflow, and the circumgalactic medium. The complex morphology and kinematics imply that the quasar resides in a very dense environment with several interacting companion galaxies within projected distances of 10--15~kpc. The high density of the environment and the large velocities of the companion galaxies suggest that this system may represent the core of a forming cluster of galaxies. The system is a good candidate for a merger of two or more dark matter halos, each with a mass of a few $10^{13}$~M$_\odot$ and traces potentially one of the densest knots at $z\sim3$.
\end{abstract}



\section{Introduction} 
\label{sec:intro}

Feedback from active galactic nuclei (AGN) is a standard ingredient in galaxy evolution models and is invoked to explain the steep decline of the galaxy mass function and to establish the black hole vs. bulge correlations \citep{fabi12, veil05, Veilleux20}. Popular evolution models predict that black holes grow initially in obscurity, deep inside a dusty galactic starburst, until a blowout of gas and dust, driven largely by the AGN, leads to a cessation of the star formation and a luminous quasar ($L_{\rm{bol}} > 10^{45}$ \ergs) becomes visible in the galactic nucleus \citep[e.g.][]{hopk08}. In many models, this process is closely linked to mergers and galaxy interactions triggering the starburst and black hole growth as this would provide a natural way to supply gas to the galactic center to trigger the starburst and the quasar. Red, dust-obscured quasars are expected to be valuable test-cases for this scenario. Indeed, evidence has been building up that red, dusty, luminous quasars at $z \sim 2-3$ are associated with particularly fast ionised outflows and are sign-posts for the most active ‘blow-out’ phase of quasar feedback \citep{zaka16b, perr19, CalistroRivera_2021, Vayner21a}. 

While this is a compelling scenario, observational support for enhanced merger activity in quasar hosts relative to inactive galaxies remains elusive. This is in part due to the glow from the quasar itself which acts as a major limiting factor and has impeded studies of the quasar hosts in the past. Even at low redshifts the connection between quasar activity and mergers is controversial \citep{wyle16a, vill17, goul18b}, and the picture is even less clear at Cosmic Noon ($z \sim 2$), when quasar activity peaked, and mergers and galaxy interactions were more common \citep{Bischetti_2021} but where observations are complicated both by cosmological surface brightness dimming and by the relative compactness of high-redshift massive galaxies. Even with the improved sensitivity and resolution of Wide Field Camera 3 (WFC3) on the \textit{Hubble Space Telescope} (\hst), disentangling the quasar emission from the faint and compact host galaxy has been challenging and results are not yet conclusive \citep{chia15, glik15, mech16, farr17, zaka19}. Possible companion galaxies and interacting galaxies are particularly difficult to confirm given the high density of foreground and background sources and the usual lack of spectroscopic information in a wider field, particularly for sources at Cosmic Noon. 

In this paper we present the first results of integral-field unit (IFU) observations of the luminous quasar SDSS J165202.64+172852.3 (J1652 hereafter) at $z \sim 3$ taken with the Near Infrared Spectrograph \citep[NIRSpec; ][]{Jakobsen_2022} aboard the James Webb Space Telescope \citep[\jwst; ][]{Gardner_2006}. J1652 belongs to the class of Extremely Red Quasars (ERQs), a population of luminous ($\ga 10^{47}$ \ergs) quasars at $z>2$ identified by their very high infrared-to-optical ratios and peculiar optical spectra \citep{ross15, hama17} using the combination of data from Wide-field Infrared Survey Explorer (WISE; \citealt{wrig10}) and from the Baryon Oscillation Spectroscopic Survey (BOSS; \citealt{daws13}) of the 3rd generation of the Sloan Digital Sky Survey (SDSS-III; \citealt{eise11}). J1652 has a $i-W3$ color of 5.4 mag, well above the ERQ selection cut-off of $i-W3 > 4.6$~mag and rest equivalent width (REW) of \civ\ of 125\AA\ (ERQ selection REW(\civ)\ $ \geq 100$~\AA). ERQs have also shown to have extreme ionized outflows \citep{zaka16b, perr19} and they are prime candidates for the early `blowout/transition phase' of quasar/galaxy evolution. J1652 is one of the most powerful obscured quasars at its epoch with near Compton-thick column densities \citep{goul18a, Ishikawa21} and is a $\sim 1.5$~mJy radio-intermediate source \citep{alex17, hwan18}. Its bolometric luminosity of $5\times 10^{47}$~\ergs\ is estimated based on the directly observed infrared flux from WISE \citep{wrig10} and bolometric corrections applicable for obscured quasars \citep{goul18a}. Ground-based near-IR IFU observations reveal broad \oiii~5007\AA\ emission (velocity width $\sim$ 3000 \kms) which is associated with a quasar-driven outflow and extends $\sim$ 3 kpc towards the South \citep{alex18, Vayner21a}. \hst\ WFC3 IR observations tracing stellar light in the rest-frame $B$-band suggest the host galaxy is very massive: $\log{M_*/M_\odot \sim 11.4-12.4}$. These images also show extended emission in the western direction, likely originating from a tidal tail \citep{zaka19}. Based on this morphology, \citet{zaka19} proposed J1652 as a candidate major merger, even though no massive companion galaxy could be identified. 

Existing observations lack the sensitivity, resolution, and contrast to reveal the true extent of the outflow and to map the environment. Focusing on signatures of the ionized gas traced by \oiii, we present new \jwst\ near-IR IFU observations which uncover the complex details of the J1652 environment (20-30~kpc) for the first time. A multi-line, in-depth analysis of the NIRSpec data for this quasar will be presented in an upcoming paper by Vayner et al. in prep. In Section \ref{sec:obs_redux}, we describe the design of the observations and the data reduction. In Section \ref{sec:analysis}, we present measurements and analysis, and we discuss our results in Section \ref{sec:conclusions}. Following a long-standing convention, emission lines are identified by their wavelength in air (e.g., \oiii$\lambda$5007\AA), but all wavelength measurements are performed on the vacuum wavelength scale. We use a $H_{0} = 70$~km~s$^{-1}$~Mpc$^{-1}$, $\Omega_m = 0.3$ and $\Omega_{\lambda} = 0.7$ cosmology throughout this paper.

\begin{figure*}
\centering\includegraphics[width=1\textwidth, trim = 0cm 0cm 0cm 0cm, clip=true]{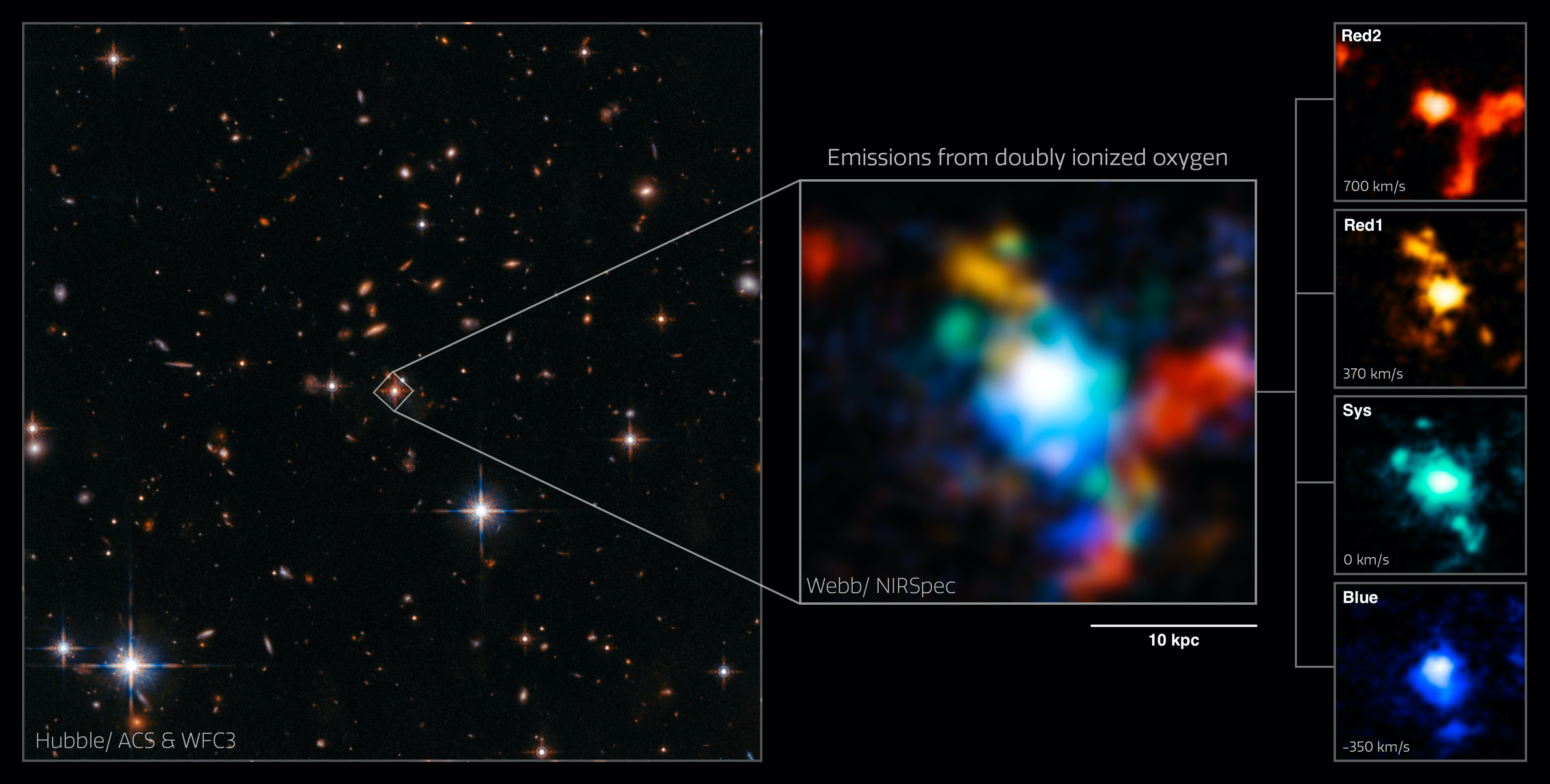} \\
\caption{Next to the HST ACS \& WFC3 wide-field composite, we show the four-color image (middle) showcasing the complex \oiii\ kinematics and morphology around the extremely red quasar J1652 at $z=2.9489$ from the JWST data. North is up and East is to the left. The quasar position is at the center of the image. The image is $\sim$ 26~kpc on the side and is composed of four narrow-band images made from the NIRSpec cube. The narrow-band images, shown individually in the right panels, are centered on the \oiii5007\AA\ emission line at velocities of -350, 0, 370, and 700~\kms\ and they are $\sim$120~\kms\ wide. The four-color image reveals the bi-conical (blue- and red (orange)-shifted emission) quasar-driven outflow in the NE-SW direction as well as a companion galaxy (Companion C1) at $\sim 700$~\kms\ in the North-East corner of the NIRSpec cube (red emission). In addition, the NIRSpec cube reveals an extended \oiii\ component at $600-800$~\kms\ (red) in the South-Western quadrant of the image. This extended component is associated with continuum stellar emission as revealed by the \hst\ data. Image Credit: ESA/Webb, NASA \& CSA, D. Wylezalek, A. Vayner \& the Q3D Team, N. Zakamska}
\label{nb_images}
\end{figure*}

\section{Observations and Data Reduction} \label{sec:obs_redux}

\subsection{Observations}

To realize \jwst's full science potential, the Space Telescope Science Institute (STScI) and the \jwst\ Advisory Committee developed the Director’s Discretionary-Early Release Science (DD-ERS) program. The ERS observations are taking place during the first 5 months of \jwst\ science operations and do not have any proprietarty period. The aim of the ERS program is for the science community to quickly learn to use its instruments and capabilities. The program `Q-3D: Imaging Spectroscopy of Quasar Hosts with \jwst\ Analyzed with a Powerful New PSF Decomposition and Spectral Analysis Package' (or Q3D for short; ID: 1335, PI: Wylezalek, Co-PIs: Veilleux, Zakamska, Software Lead: Rupke) was selected as one of the 13 ERS programs and is one of two science programs in the category `Massive Black Holes and Their Host Galaxies' \footnote{\href{https://q3d.github.io}{https://q3d.github.io}} \citep{Wylezalek_2017}. Observations for Q3D were designed with the help of the \jwst\ User Documentation \citep{Jdox}.

The first target of the Q3D program, SDSSJ1652, was observed on 2022-07-15 and 2022-07-16 by \jwst\ using the NIRSpec Instrument in IFU mode \citep{Boeker_2022, Jakobsen_2022}. Some of the data presented in this paper were obtained from the Mikulski Archive for Space Telescopes (MAST) at the Space Telescope Science Institute. The specific observations analyzed can be accessed via \dataset[10.17909/qn07-rt28]{https://doi.org/10.17909/qn07-rt28}. The NIRSpec field-of-view in IFU mode is $\sim 3\arcsec \times 3 \arcsec$ ($\sim 26 \times 26$~kpc). We used the filter/grating combination F170LP/G235H, with corresponding wavelength coverage $1.70-3.15~\mu$m (corresponding to $\sim 0.43-0.79$~$\mu$m at the redshift of J1652). The grating has a near-constant dispersion $\Delta\lambda = 8.7\times10^{-4}~\mu$m, corresponding to velocity resolution $85-150$ \kms. This allows us to spectrally resolve multiple kinematic components in the emission lines, which have typical velocity widths of several hundred \kms. To improve the spatial sampling---in order to accurately measure and characterize the point spread function (PSF)---we used a 9-point small cycling dither pattern with 25 groups and 1 integration per position. To account for light leaking through the closed micro-shutter array (MSA), as well as light from failed open shutters, we took one leakage exposure at the first dither position. We used the NRSIRS2 readout mode, which improves signal-to-noise but reduces data volume compared to the NRSIR2RAPID mode. No pointing verification image was taken. The total integration time was 4.6 hours on target and 0.5 hours for the leakage exposure.

\subsection{Data Reduction}
\label{subsec:Data_reduction}

Reduced level 2 and 3 data products available on the MAST archive used an outdated version of the pipeline with pre-flight calibration files. Therefore, to improve the data reduction quality, we ran the latest pipeline on the uncalibrated level 1 data downloaded from the archive.

Data reduction was made with the \jwst\ Calibration pipeline version 1.6.2 using CRDS version ``11.16.8' and context file ``jwst 0945.pmap'. The first stage of the pipeline, \verb|Detector1Pipeline|, performs standard infrared detector reduction steps such as dark current subtraction, fitting ramps of non-destructive group readouts, combining groups and integrations, data quality flagging, cosmic ray removal, bias subtraction, linearity, and persistence correction. 

Afterward, we ran \verb|Spec2Pipeline|, which assigns a world coordinate system to each frame, applies flat field correction, flux calibration, and extracts the 2D spectra into a 3D data cube using the \verb|cube build| routine. At present the pipeline still uses pre-flight flux calibration files, however, we do not require absolute flux calibration for the analysis presented in this paper. For NIRSpec, additional steps are taken to flag pixels affected by open MSA shutters. At this point, we skipped the imprint subtraction step due to increased spatial variation in the background across many spectral channels. Due to known issues with the outlier detection step in the \verb|Spec3Pipeline| (\jwst\ Help Desk, priv. communication), we opted to use the \verb|Montage| package\footnote{\href{http://montage.ipac.caltech.edu}{http://montage.ipac.caltech.edu}} to combine the different dither positions into a single data cube using their drizzle algorithm. The dither positions are combined onto a common grid with a spatial pixel size of 0.05\arcsec.


\subsection{PSF Subtraction}

To remove the bright, spatially unresolved quasar emission, we employ the method outlined in \citet{Vayner_2016,Vayner2021c}. We are developing a dedicated software package for PSF subtraction, \texttt{q3dfit} (Rupke et al. 2022, in prep.), which we will apply to these data in future work. For this initial analysis, we construct a PSF model with a spatial distribution that is wavelength-independent using a small wavelength range near \oiii. We use the blueshifted wing of the \hb\ line, selecting velocity channels offset by $<-3000$ \kms\ from the quasar redshift to avoid extended \hb\ emission from the quasar host galaxy. In particular, we use the $[1.898\micron ,~1.900 \micron]$ range, median combining five spectral channels. Next, we apply a 5$\sigma$ signal-to-noise ratio cut on the resulting image, leaving behind only flux from the PSF structure that has a full width at half maximum of $\sim 200$~mas. Finally, we normalize the PSF image by the peak flux. To remove the PSF from the data in each spectral channel in the range $[1.7002 \micron,~2.42945 \micron]$, we scale the flux of the PSF image by the peak flux in that channel. Finally, we perform additional processing to mitigate issues caused by spatial under-sampling of the {\it JWST} PSF by the NIRSpec IFU at these wavelengths. We will discuss these additional steps in a forthcoming paper (Vayner et al. in prep.) 

\subsection{Astrometric correction}
\label{subsec:astrometry}

The astrometry of the cube obtained by the JWST Calibration pipeline (see Section \ref{subsec:Data_reduction}) features a minor offset with respect to the ground-based data. We therefore use the supporting PSF-subtracted \hst\ WFC3 F160W image from \citet{zaka19} to align the astrometry based on SDSSJ1652's companion galaxy in the North-East (C1 in Section \ref{sec:analysis} below). In both the \hst\ image and an \ha\ narrow-band image created from the NIRSpec cube, we fit the brightness profile with a 2D Sérsic model to determine the location of the peak flux. We then overlay the peak positions on top of each other, retrieving an offset of $\Delta \rm R.A. =0.040\arcsec$ and $\Delta \rm Dec. =1.021\arcsec$.

\section{Analysis}
\label{sec:analysis}

\subsection{Morphology}
\label{sec:Morphology}

The complex morphology and spatial extent of the \oiii\ emission are clearly revealed in a four-color image (middle panel of Figure \ref{nb_images}). The individual narrow-band images (shown separately in the right panels) that make up the color-composite are obtained by collapsing the NIRSpec cube across four separate velocity ranges at [$-370, -250$], [$-20, 100$], [$350, 470$], and [$700, 820$]~\kms. We refer to the individual resulting narrow-band images as Blue, Sys, Red1 and Red2 from here on. These velocity ranges were chosen as best representations of the multiple kinematic components observed in and around J1652 (see also Section \ref{sec:Kinematics}). We use the updated redshift of $z = 2.9489$ (Section \ref{sec:Kinematics}) to shift the wavelength axis to rest-frame. All velocities are measured relative to this frame. Each narrow-band image is two spectral bins wide corresponding to a spectral width of $\sim 120$~\kms. Additionally, we perform a continuum subtraction using the average signal in a line-free spectral range blue- and redward of the \oiii/H$\beta$ line complex, respectively, at $4740-4790$ \AA\ and $5080-5120$ \AA\ in the rest frame. 

\begin{figure*}
\centering
\includegraphics[width=0.95\textwidth, trim = 1cm 1cm 3cm 3cm, clip=true]{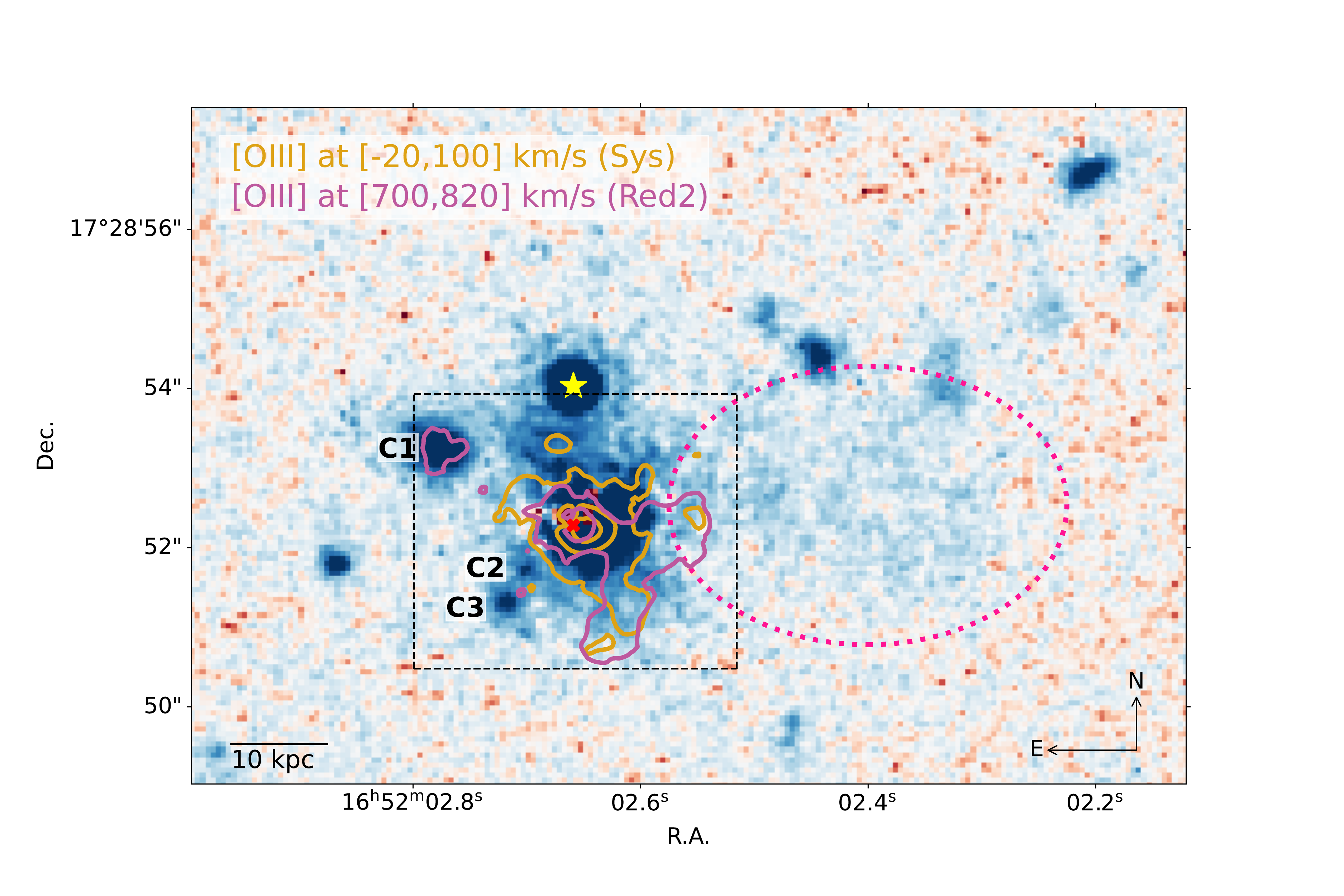}
\caption{We show the PSF-subtracted \hst\ WFC3 F160W image adopting a colour scheme that highlights faint features showing the negative, zero, and positive values in blue, white, and red, respectively \citep[see also][]{zaka19}. The extended stellar light emission is clearly visible towards the West and we mark it by the pink ellipse. We also show the NIRSpec contours of the narrow-band images Sys and Red2 and show the NIRSpec footprint by the black dashed box. The NIRSpec-detected \oiii\ emission, extending into the same direction as the \hst\ extended emission (marked with the pink ellipse) confirms that the extended emission feature is associated with the J1652 system. With the NIRSpec data, we furthermore confirm that three \hst\ continuum sources are companion sources (C1, C2, and C3). The foreground star in the field is marked by the yellow star symbol and the position of the quasar J1652 by the red cross.}
\label{overlays}
\end{figure*}

We detect \oiii\ emission across the entire field of view, which is 26~kpc (in physical units) on each side. Multiple narrow kinematic components are apparent in the North (e.g. in Sys at $\sim 0$~\kms) and South (e.g. in Sys and Red2 at $\sim 0$~\kms\ and $\sim 700$~\kms, respectively). In particular, the emission in the North-Eastern corner of the Red2 narrow-band image is associated with a companion galaxy (Companion C1, see Figure \ref{overlays}) that had been detected as a continuum source in \hst imaging. While it is not apparent in the color-composite, we also detect narrow line emission as well as continuum emission associated with two additional \hst-detected companions C2 and C3 (see Figures \ref{overlays} and \ref{kinematics}). In Figure \ref{overlays}, we show the PSF-subtracted \hst\ WFC3 F160W image originally published in \citet{zaka19} together with the contours from the \oiii\ NIRSpec Sys and Red2 narrow-band images. The black dashed rectangle marks the footprint of the NIRSpec observation while the object highlighted with the star symbol marks a foreground star in the field. C1, C2 and C3 were clearly detected in the \hst\ image, but their physical association with the J1652 system was unknown since no spectroscopic information was available. The NIRSpec-derived redshifts now confirm that they are companion galaxies of J1652. 

While \oii\  3727 \AA\ is within the F160W filter at the redshift of our target, its rest equivalent width rarely exceeds 100 \AA\ \citep{Reddy_2018} and therefore in a 3000 \AA-wide filter it accounts for no more than 13\% of the measured fluxes. As a result, the \hst\ image traces line-free stellar light. The best Sersic fit to the host of J1652 yields rest-frame $B-$band luminosity of $L_B=10^{12.0}L_{\odot}$ \citep{zaka19}, which is significantly above the break of the galaxy luminosity function at $10^{10.6} L_{\odot}$ at a similar redshift \citep{gial05}. We now perform aperture photometry on the companions in the \hst\ data and find that their F160W fluxes are 12.3, 2.5 and 3.2\%, respectively, of the J1652 host. The mass-to-light ratio of high-redshift galaxies in the $B-$band is very uncertain \citep{zaka19}, but with an estimated stellar mass for J1652 in the range $10^{11.4-12.4}M_{\odot}$, the companions' stellar masses are in the range of $10^{9.8-11.5}M_{\odot}$. 

The \hst\ image also clearly shows extended emission in the Western direction (marked by the pink ellipse). \citet{zaka19} speculated that this emission might be a stellar tidal feature and they identified J1652 as a candidate major merger. With the NIRSpec data, we now confirm that the emission is indeed associated with the J1652 system. In Section \ref{sec:conclusions} we discuss the nature of this emission.

\subsection{Kinematics}
\label{sec:Kinematics}

\begin{figure*}
\includegraphics[width=1\textwidth, trim = 7cm 5cm 7cm 5cm, clip=true]{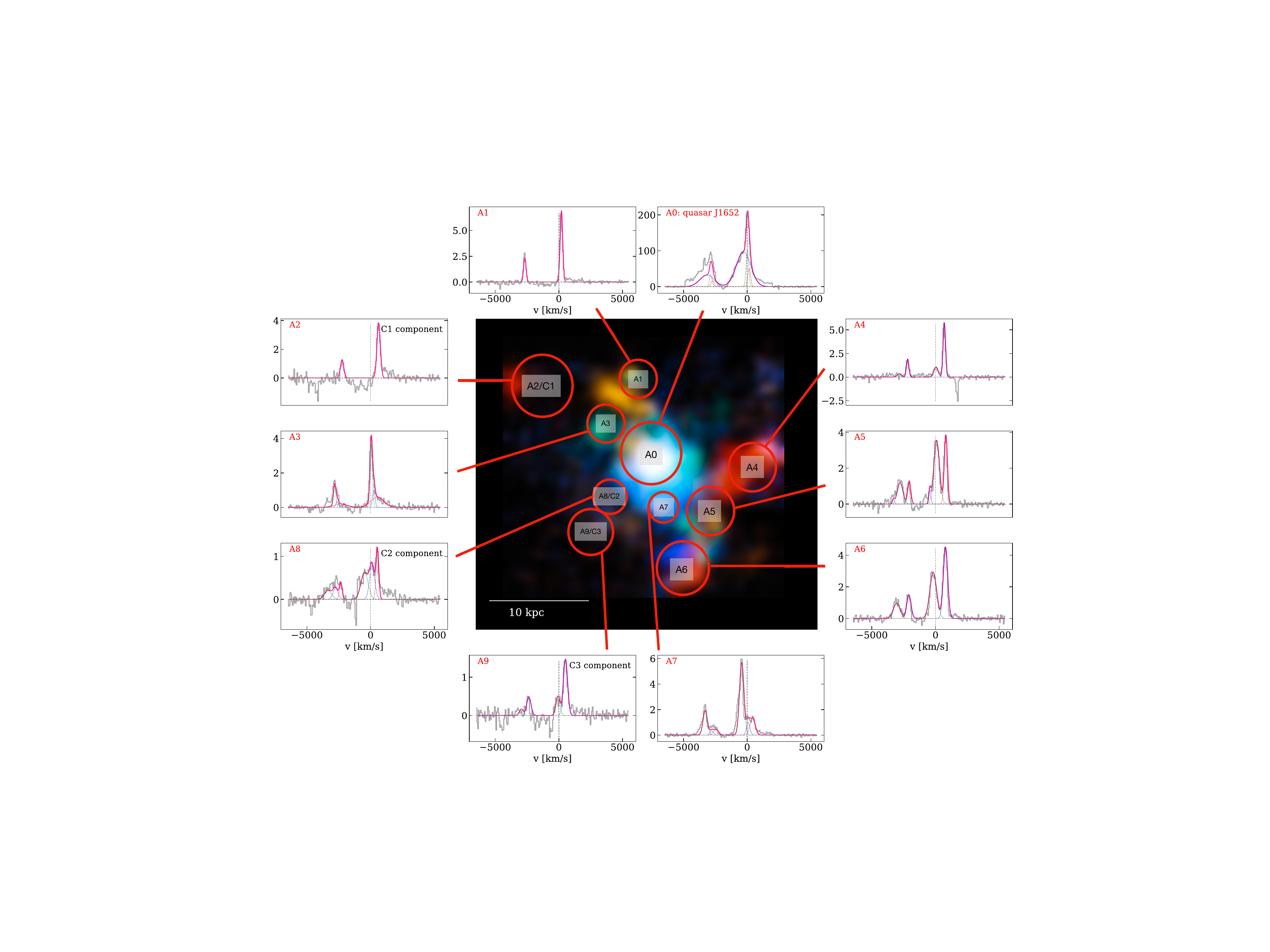}
\caption{NIRSpec [O~III] color composite (from Figure \ref{nb_images}) along with the NIRSpec spectra (in dimensionless $F_{\lambda}$ units) zooming into the [O~III] doublet at different positions in the NIRSpec field of view. The spectra reveal spatially well-defined individual kinematic components at velocities ranging between $v = -450$~\kms\ to $v = 820$~\kms\ and velocity dispersions between $\sigma = 50-500$~\kms. The fits are performed by using up to three Gaussian components and the fit results are presented in Table \ref{fit_results}. The companion galaxies C1, C2 and C3 (see Figure \ref{overlays}) are spectroscopically confirmed in Apertures A2, A8 and A9, respectively, at velocity offsets of $\sim 600$~\kms\ with respect to the quasar host [O~III] emission. We also detect extended [O~III] emission associated with the extended stellar feature seen in the \hst\ images (e.g. A4, A5, A6) at velocities of $\sim 700-800$~\kms. Additionally, we confirm the blue-shifted component of the quasar-driven outflow (e.g. in A0 and A7) and newly identify the likely red-shifted counter-part in the North (e.g. detected in A3). The outflow axis well matches previous constraints of the projected quasar illumination direction from polarization measurements of this source \citep{alex18}. Image credit of central color composite: ESA/Webb, NASA \& CSA, D. Wylezalek, A. Vayner \& the Q3D Team}
\label{kinematics}
\end{figure*}

Motivated by the morphology of the different kinematic components apparent in the narrow-band images, we extract and fit the \oiii\ line in ten apertures. Figure \ref{kinematics} shows the position and spectra of the extraction apertures, referred to as A0 - A9. A0 corresponds to the position of the quasar J1652 itself. Due to known remaining flux calibration issues with NIRSpec data cubes, we show all spectra in dimensionless $F_{\lambda}$-units. Based on the fit to A0 which, in addition to a broad blue-shifted component related to a quasar-driven outflow \citep{Vayner21a}, traces the narrow \oiii\ emission associated with the quasar host, we update the redshift estimate to $z = 2.9489$ and are reporting velocities relative to this frame. The spectral fits, performed by using up to three Gaussians, reveal spatially well-defined individual kinematic components at velocities ranging between $v = -450$~\kms\ to $v = 820$~\kms\ and velocity dispersions between $\sigma = 90-500$~\kms (corrected for instrumental resolution effects). The results of the individual fits are reported in Table \ref{fit_results}. 

The narrow component associated with the host of J1652 at $v \sim 0$~\kms\ is apparent in all apertures apart from A1 and A2. This is indicative of a large systemic galaxy component with extents of $\sim 10$~kpc from the galaxy nucleus, roughly consistent with the stellar component of J1652 after PSF subtraction (see Fig. \ref{overlays}). The gas distribution at $v \sim 0$ \kms\ appears clumpy (see narrow-band image S at $v=[-20,100]$~\kms\ in Figure \ref{nb_images}). We confirm the ionised gas outflow towards the South-West of the quasar host (A0, A7, broad component with $v = -300$~\kms\ and $\sigma = 570$~\kms; \citealt{Vayner21a}) and newly detect the red-shifted counter-part of the outflow in the North-East which is receding (A3, broad component with $v = 450$~\kms\ and $\sigma = 500$~\kms). We note that the known presence of dust in this source may impact the morphology and measured extents of the observed structures. For example, the Northern part of the outflow was previously undetected in the shallower [OIII] ground-based NIFS observations. The back cone is likely partially extincted by dust \citep{Vayner21a} and the [OIII] line strength was below the flux sensitivity of NIFS which is about 1 dex shallower in surface brightness compared to the new NIRSpec observations.

Additionally, we find emission at $v = 640$~\kms\ in A2 and at $v = 510-530$~\kms\ in A8 and A9 that are respectively associated with continuum sources C1, C2, and C3 (Figure \ref{overlays}). The velocity dispersion of 130$-$160~\kms\ is in the normal range for undisturbed galaxies. We therefore confirm the presence of at least three companion galaxies of J1652 at projected distances of $\sim 5-10$~kpc. \cite{zaka19} also reported extended emission detected in \hst\ WFC3 imaging in the western direction likely originating from a stellar tidal feature. With NIRSpec, we find ionised gas at $v \sim 700-800$~\kms\ associated with that feature extending across the entire South-Western quadrant of the cube (A4, A5, A6, A7 and A8). The gas is kinematically cold with $\sigma \sim 100 - 200$~\kms.

\begin{table*}[t]
\caption{Results from the multi-component Gaussian fitting in Apertures 0-9 as shown in Figure \ref{kinematics}. The apertures in this table are ordered roughly by velocity offset of the dominating component.}
\label{fit_results}
\begin{tabular}{c|cc|cc|cc|cc}
\hline
 & $v_{\rm blue}$ & $\sigma_{v_{\rm blue}}$ & $v_{\rm sys}$ & $\sigma_{v_{\rm sys}}$ & $v_{\rm red1}$ & $\sigma_{v_{\rm red1}}$ & $v_{\rm red2}$ & $\sigma_{v_{\rm red2}}$\\
 & km~s$^{-1}$ & km~s$^{-1}$ & km~s$^{-1}$  & km~s$^{-1}$& km~s$^{-1}$ & km~s$^{-1}$ & km~s$^{-1}$ & km~s$^{-1}$ \\
\hline
A0\tablenotemark{a} & -289 & 548 & 19 & 101 &  156 & 113 & -- & -- \\
A7 &-427 & 111 & 81 & 18 & 430 & 147 & -- & -- \\
A8\tablenotemark{b} &-475 & 250 & 125 & 148 & 546 & 107 & -- & -- \\
A5 & -437 & 81 & 92 & 177 & -- & -- & 819 & 123 \\
A6 &-177 & 274 & -- & -- & -- & -- & 788 & 109 \\
A3 & -- & -- & 53 & 94 & 251 & 102 & 526 & 481 \\
A4 & -- & -- & 42 & 99 & -- & -- & 696 & 98\\
A1 & -- & -- &--&-- & 205 & 100&--&--\\
A2\tablenotemark{c} & -- & -- & --  &--&--&--& 640 & 134\\
A9\tablenotemark{d} & -- & -- & -68 & 83 & 520 & 100 &--&--\\
\hline
\end{tabular}
\tablenotetext{a}{Aperture centered on quasar position.}
\tablenotetext{b}{Corresponds to companion galaxy C2.}
\tablenotetext{c}{Emission at 500~km/s is associated with companion galaxy C1.}
\tablenotetext{d}{Corresponds to companion galaxy C3.}
\end{table*}

\section{Discussion and Summary}
\label{sec:conclusions}

In this paper, we explore the wealth and complexity of the \oiii\ ionized gas emission of the extremely red quasar J1652 using \jwst\ NIRSpec observations in light of previous ground-based near-IR observations and \hst\ imaging. J1652 is a very luminous, extremely red quasar with $L_{\rm{bol}} \sim 5 \times 10^{47}$~\ergs\ \citep{goul18a} that drives strong ionised gas outflows on scales of 1-15~kpc. Extended broad, blue-shifted emission towards the south-west was previously detected using ground-based, laser-guided adaptive optics NIFS observations \citep{Vayner21a}. Using \jwst\ NIRspec observations we confirm the blue-shifted outflow and additionally detect the corresponding red-shifted part of the outflow towards the North-East. The orientation agrees with the previous \oiii\ measurements, as well as with imaging and polarimetric observations which trace the geometry of quasar light scattered into our line of sight \citep{alex18, Vayner21a}. This suggests that, despite the complexity of the galactic and circumgalactic environment, the illuminated part of the outflow is bipolar \citep{wyle16a} and therefore the obscuration of the nucleus is well organized in an axisymmetric structure \citep{anto93}. A full multi-line, in-depth analysis of the NIRSpec data for this quasar will be presented by Vayner et al. in prep. and will be complemented by upcoming MIRI MRS observations. 

The picture that emerges in this first look at the NIRSpec data is one that we did not anticipate. \hst\ imaging had already suggested that J1652 was undergoing a major merger due to a large, but faint tidal feature. Obscured quasars undergoing major mergers in the nearby universe sometimes host large-scale, complex \oiii\ nebulae at 10--20~kpc scales \citep{rodr2014, Leung_2021}. Larger, 100~kpc nebulae may be common around $z \sim 0.5-1$ Eddington-limited quasars \citep{villar_2010, Johnson_2018, Helton_2021}. Similar to our situation, the velocity dispersions (up to $\sim 1400$~km/s) of the host groups at these redshifts seem to arise from a combination of outflows and stripped gas and imply dark matter halo masses of a few $\times 10^{13}M_{\odot}$. At $z < 1$, such signatures can arise in galaxy-group environments but the structures are unlikely to collapse into massive clusters by $z=0$.

The NIRSpec observations reveal that the luminous red quasar J1652 at $z = 2.9489$ is in fact located in a very dense environment with several interacting companion galaxies within a projected distance of $12$~kpc of the quasar with velocity offsets of the companions of 500-700~\kms. As we lay out in the following paragraphs, these observations suggest that this system may represent the core of a (proto-)cluster of galaxies, potentially one of the densest knots at its redshift, with the potential to collapse into a Coma-like structure at low redshift. 

Distant luminous quasars hosted by massive galaxies, for example high-redshift radio galaxies, have indeed been shown to be excellent tracers of galaxy overdensities up to the highest redshifts \citep{Wylezalek_2013, Wylezalek_2014, Hatch_2014}. Quasars hosted by galaxies with stellar masses of $>10^{10.5}$~M$_{\odot}$ reside in dark matter haloes of $>10^{12}$~M$_{\odot}$ \citep{Shen07, Behroozi_2010, Hartley_2013}. Such massive haloes at $z > 1.5$ typically grow into cluster-mass structures by today \citep{Chiang_2013}. 

J1652 is hosted by a particularly massive galaxy: its rest-frame $B-$band luminosity is the highest among the ERQs investigated by \citet{zaka19} implying a host galaxy stellar mass of $\log{M_*/M_{\odot} \sim 11.4-12.4}$. We now find that its immediate environment probed by NIRSpec is very reminiscent of the cores of some of the most distant confirmed protoclusters. Two new pieces of evidence in our NIRSpec observations suggest that J1652 is in a particularly massive overdensity. 

One is the number density of companion galaxies. Three companions with estimated stellar masses $>10^{9.8}M_{\odot}$ are confirmed to be at the redshift of J1652 by NIRSpec observations, and multiple other objects in the field seen in the wider field \hst\ image are also candidate companions. Assuming that the three confirmed companions are within a physical distance of $\sim 100$ kpc from J1652 along the line of sight, we derive a nominal density of galaxies per unit co-moving volume of five orders of magnitude above the field value \citep{Davidzon_2017}. Even in comparison with known overdense environments and protoclusters around high-redshift radio galaxies \citep{Wylezalek_2014}, the nominal density within the NIRSpec field of view is at least two orders of magnitude higher. Thus with NIRSpec we are likely probing the dense central core of a massive dark matter halo, only 12~kpc from the center, whereas in the protoclusters of \citet{Wylezalek_2014}, the galaxy density is averaged over a $\sim$1 Mpc field of view. The centers of local redshift galaxy clusters are also hosts to cD galaxies, the most massive known galaxies in the Universe. Since these galaxies are only found at the centers of galaxy clusters, the cluster environment is likely tightly linked to their formation. Given the galaxy density in the here probed projected volume, J1652 may be a viable cD galaxy progenitor.

The second piece of evidence is the large velocity spread of the companions as well as the gas clumps detected in the NIRSpec field. It is difficult to achieve a velocity range of $\sim 1000$ \kms\ near the center of a halo unless its mass is well above $10^{14} M_{\odot}$ \citep{Hearin17}, which would be implausible at the redshift of our target \citep{Lukic_2009}. Motions of the cluster core relative to the bulk can increase the apparent velocity range \citep{Behroozi_2013}, and so can preferential selection of star-forming galaxies \citep{Wu_2013} likely in observations focused on line emission. Furthermore, it would not be surprising if a massive halo at such high redshift is not yet virialized, and with its prominent stellar tidal tails J1652 may be a good candidate for a merger of two or more dark matter halos, each with a mass of a few $\times 10^{13}M_{\odot}$. In this case, the apparent velocity spread may be increased by a factor of $2-3$ compared to the virialized value \citep{Kuiper_2011}, which would explain the very high range of velocities of the companions with respect to J1652. 

At high redshift, there are now observations of a handful of other protoclusters sharing three key properties -- large stellar mass of the central galaxy, large number of nearby companions and their high velocity dispersion -- with J1652 \citep{Shi_2021, Ginolfi_2022}. For example, a powerful radio-loud AGN PKS 1138-262 at $z \sim 2.2$ is hosted by a massive galaxy with $M_*\sim 10^{12}$~M$_{\odot}$ \citep{Seymour_2007, Hatch_2009}, also known as the Spiderweb galaxy. Its environment is the prime example of a forming galaxy cluster. Deep \hst\ imaging shows tens of candidate satellite galaxies \citep{Miley_2006} within $\sim 100$ kpc. At least 11 are confirmed to be at the redshift of the protocluster \citep{Kuiper_2011}, implying a galaxy overdensity of $\sim$ 200 in the central $\sim 60$ kpc relative to the field. In direct relevance to our observation of J1652 is the small-scale environment around the Spiderweb galaxy. \citet{Kuiper_2011} confirm at least three companion galaxies within 20~kpc of the Spiderweb galaxy with narrow emission lines, interpreting their velocity offsets of $400-1100$~\kms\ as gravitational motion within the massive halo of the protocluster. On the basis of the properties of the core and of the larger-scale environment (where kinematic data hint at a bimodal distribution of velocities), as well as using comparisons with simulations \citep{Springel_2006, Delucia_2007}, \citet{Kuiper_2011} conclude that the Spiderweb protocluster is likely an ongoing merger of two galaxy groups with dark matter halo masses of several $\times 10^{13}$~M$_{\odot}$. 

Another example is a $z = 4.3$ protocluster in the field of the sub-mm source SPT2349-56 \citep{Miller_2018}. Using deep ALMA spectral imaging, multiple cluster members within 15~kpc are confirmed with velocity offsets of up to 700~\kms. This cluster is different from the Spiderweb cluster in that the observed kinematics and spatial configuration of the galaxies suggest that the structure more likely represents a single gravitationally bound halo rather than multiple groups. The velocity dispersion of $\sim 410$~\kms\ of the galaxy velocities and other arguments suggest that this system represents the core of a cluster of galaxies that is at an advanced stage of formation at $z = 4.3$.

Future spectroscopic observations of the galaxies in the larger field around J1652 will clarify the structure of its large-scale environment. There has been a long-standing discussion whether a (proto-)cluster environment is a necessary condition to trigger rare, jet-dominated radio galaxies at high $z$. If so, it would explain why radio galaxies are excellent tracers of protoclusters \citep{Hatch_2014}. Similarly, one may speculate whether a proto-cluster (merging) environment is necessary to explain the aggregate properties of the rare and extreme ERQ population. This may be tested by analyzing the large-scale environments of matched samples of ERQs and blue quasars. However, this goes beyond what is possible with currently available data sets.

Our new NIRSpec observations presented here in concert with the \hst\ imaging suggest that the 20~kpc-scale environment of the powerful, extremely red quasar J1652 is similar in terms of galaxy density and kinematics to what is observed in some of the densest knots in the high-redshift Universe \citep[e.g.][]{Kuiper_2011}. 
While it is difficult to predict the halo masses that such a structure would evolve into by $z=0$ owing to the large halo-to-halo variations in dark matter halo-growth histories \citep{Chiang_2013}, the observations are consistent with J1652 residing in a dark matter halo of $\sim 10^{13}$~M$_{\odot}$ which may be involved in a merger of two or more dark matter halos. Such systems are expected be the progenitors of galaxy cluster halos with $M > 10^{15}$~M$_{\odot}$ at $z=0$.




\begin{acknowledgments}

D.W. and C.B. acknowledge support through an Emmy Noether Grant of the German Research Foundation, a stipend by the Daimler and Benz Foundation and a Verbundforschung grant by the German Space Agency. A.V., D.S.N.R., N.L.Z., and S.V. are supported by NASA through STScI grant JWST-ERS-01335. N.L.Z further acknowledges support by the Institute for Advanced Study through J. Robbert Oppenheimer Visiting Professorship and the Bershadsky Fund. J.B.-B. acknowledges support from the grant IA- 101522 (DGAPA-PAPIIT, UNAM) and funding from the CONACYT grant CF19-39578.

This research made use of Montage. It is funded by the National Science Foundation under Grant Number ACI-1440620, and was previously funded by the National Aeronautics and Space Administration's Earth Science Technology Office, Computation Technologies Project, under Cooperative Agreement Number NCC5-626 between NASA and the California Institute of Technology.

\end{acknowledgments}

%

\vspace{5mm}
\facilities{JWST(NIRSpec), HST(WFC3) }


\software{astropy \citep{Astropy2013, Astropy2018},  Montage \citep{Cigan_2019}}





\bibliography{bib}{}
\bibliographystyle{aasjournal}



\end{document}